\documentclass{aastex62}
\usepackage{rotating}
\usepackage{amsmath}
\usepackage{graphicx}
\usepackage{enumerate}
\newcommand{\hhwang}[1]{{\color{black} #1}}

\renewcommand{\vec}[1]{\boldsymbol{#1}}

\received{}
\revised{}
\accepted{}
\submitjournal{ApJS}

\shorttitle{Self-gravitational Force Calculation on Nested Grids}
\shortauthors{Wang et al.}

\begin{document}

\title{
Self-gravitational Force Calculation of Second Order Accuracy Using Multigrid Method on Nested Grids\footnote{yen@math.fju.edu.tw}}

\correspondingauthor{Chien-Chang Yen}
\email{yen@math.fju.edu.tw}

\author[0000-0002-0786-7307]{Hsiang-Hsu Wang}
\affiliation{
Department of Physics\\
The Chinese University of Hong Kong\\
Shatin, New Territory, Hong Kong, People’s Republic of China}

\author{Chien-Chang Yen}
\affiliation{Department of Mathematics \\
Fu Jen Catholic University\\
New Taipei City, Taiwan}



\begin{abstract}
We present a simple and effective multigrid-based Poisson solver of second-order accuracy in both gravitational potential and forces in terms of the one, two and infinity norms. The method is especially suitable for numerical simulations using nested mesh refinement. The Poisson equation is solved from coarse to fine levels using a one-way interface scheme.  We introduce anti-symmetrically linear interpolation for evaluating the boundary conditions across the multigrid hierarchy. The spurious forces commonly observed at the interfaces between refinement levels are effectively suppressed. We validate the method using two- and three-dimensional density-force pairs that are sufficiently smooth for probing the order of accuracy. 
\end{abstract}



\section{Introduction} 
Self-gravitational forces are responsible for the formation of objects and structures at all scales in the Universe. The calculation of self-gravity is usually associated with the Poisson equation:
\begin{eqnarray}
\nabla^2 \Phi = 4\pi G \rho,
\end{eqnarray}
where $\Phi$ is the gravitational potential, $G$ the gravitational constant and $\rho$ the volume density. Once the potential, $\Phi$, is solved, the self-gravitational forces are then obtained through a relation $F = -\nabla \Phi$. However, given a density distribution, solving the Poisson equation is not an obvious task due to the long-range nature of gravitational forces. 

Studying structure formations, such as galaxy formation, star formation, planet formation, etc., usually involves complicated physics working at different dynamical scales. For a grid-based hydrodynamic/magnetohydrodynamic (HD/MHD) code, the technique of mesh refinement is usually among the development list for addressing multi-scale problems. Although numerical schemes of high-order accuracy have been well-developed for hyperbolic systems within the framework of mesh refinement \citep{ber84, ber89, dez93, qui91}, a Poisson solver of compatible accuracy is still under development. 

Multigrid methods have a long and fruitful history. The methods employ a multi-scale hierarchical mesh structure with decreasing spatial resolution to efficiently damp the errors at large and small scales. The theory behind the methods is mathematically well-developed. For interested readers, we refer to the books by \citet{wes04} and \citet{bri00}. Relaxation methods accelerated by multigrid techniques have been successfully applied for solving the Poisson equation in computational astrophysics. This combination makes a multigrid relaxation method easy to implement and yet enjoying a linear computational complexity. It is also relatively straightforward to deploy the methods on massively parallelized computing resources, i.e., computer/GPU clusters \citep{sch18}.

\citet{Ric08} developed an improved multigrid Poisson solver for patch-based adaptive meshes based on the work of \citet{hua00}. The accuracy of this method degrades around the zones adjacent to the interfaces between levels. When the calculation domain is partially refined, the potential calculated in this way loses the desirable second-order accuracy in terms of $L^2$ error norm. The accuracy of the corresponding gravitational forces, which are derived from the gradient of the potential, cannot be better than the potential. \citet{gui11} proposed a simpler but improved multigrid scheme for solving the Poisson equation with arbitrary domain boundaries (see also \citet{gib02}). Although their calculated potential can reach second-order accuracy in terms of $L^{\infty}$ error norm in the context of adaptive mesh refinement (AMR), the corresponding forces are still suffering spurious forces at interfaces between levels. This then degrades the accuracy of the calculated forces to the first-order accuracy in terms of $L^{\infty}$ error norm. 

It is desirable to reduce the spurious forces at fine-coarse interfaces as much as possible to avoid the accumulating effects such as deflection or artificial "heating" when particles/fluids are passing through a jump in spatial resolution. It is also important to reduce the effects of impacts of "squareness" due to the use of Cartesian coordinates. Based on those previous efforts, in this work, we further improve the multigrid method for the Poisson equation applicable to nested mesh refinements. By comparing the numerical results with analytic solutions, we demonstrate the proposed method is of second-order accuracy in gravitational forces in terms of $L^{\infty}$ error norm. That is the spurious forces near coarse-fine interfaces are largely suppressed. 

The paper is organized as follows. In \S~2, we detail the algorithm of the improved multigrid method. In \S~3, we implement the proposed algorithm and conduct numerical tests for two and three-dimensional problems. The numerical results are compared with analytic solutions to validate the method. We conclude this work with a brief summary and discussions in \S~4.

\section{A multigrid algorithm for nested mesh refinement} \label{sec:method}
We follow the multigrid procedure described in \citet{gui11}, except for the treatment on the ghost zones surrounding refined levels and across the multigrid hierarchy. For completeness, we detail the entire numerical procedure in the following subsections. 
\subsection{The structure of nested mesh refinement}
We restrict the discussion to Cartesian coordinates  and describe a numerical computational domain using $\Omega^0=[\omega_L, \omega_R] \times [\omega_F, \omega_B]$ for two-dimensional problems and $[\omega_L, \omega_R]\times[\omega_F, \omega_B]\times[\omega_D, \omega_T]$ for three-dimensional problems, where the superscript 0 is used to denote the base (coarsest) level of nested mesh. The domain of the refined mesh of level $\ell$, $\Omega^{\ell}$, should satisfy:
\begin{eqnarray*}
\Omega^{\ell} \subset \Omega^{\ell-1},
\end{eqnarray*}
for $\ell =1, 2, ..., \ell_{\rm max}$, where $\ell_{\rm max}$ is a positive integer representing the highest (finest) level of mesh refinement. A simple nested  mesh hierarchy between two successive levels is shown in  Fig.~\ref{fig:MGgrid}a. 

The calculation domain of a refined patch of level $\ell$ is enclosed by the thick-black rectangle and denoted as $\Omega^{\ell}$,  while the domain of the coarse level $\ell-1$ is enclosed by the thick-blue rectangle and denoted as $\Omega^{\ell -1 }$. Each calculation domain is uniformly subdivided into cells, with $\Delta x^{\ell}=(\omega_R-\omega_L)/2^{\ell}$, $\Delta y^{\ell}=(\omega_B-\omega_F)/2^{\ell}$ and $\Delta z^{\ell}=(\omega_T-\omega_D)/2^{\ell}$ being the cell sizes of the refined level $\ell$ in $x$, $y$ and $z$ directions, respectively. Hereafter, we further assume $\Delta x^{\ell} = \Delta y^{\ell} = \Delta z^{\ell}$ so that each cell is a square (2D) or a cube (3D). The ratio of cell size between levels $\Omega^{\ell-1}$ and $\Omega^{\ell}$ is 2, i.e, $\Delta x^{\ell-1}/\Delta x^{\ell}= 2$, $\Delta y^{\ell-1}/\Delta y^{\ell}=2$ and $\Delta z^{\ell-1}/\Delta z^{\ell}=2$. A patch of level $\ell$ is entirely immersed in a coarse patch of level $\ell-1$, i.e., the level difference between patches shall be no larger than 1. Within each calculation domain, the discretized density distribution is given at cell centers.  

For simplicity, we solve the Poisson equation on a level-by-level basis, from a level $\ell-1$ to a level $\ell$, based on the so-called one-way scheme. That is, we solve the gravitational potential for the level $\ell-1$ first, and the result is used to interpolate the Dirichlet boundary values required for the refined level $\ell$. The green layer shown in Fig.~\ref{fig:MGgrid}b represents the buffer zone, denoted as $\Omega^{\ell}_b$,  surrounding the patch $\Omega^{\ell}$.  The domain $\Omega^{\ell}\cup\Omega^{\ell}_b$ then forms the calculation domain where we apply the multigrid method as shown in Fig.~\ref{fig:MGgrid}c. The buffer zone, $\Omega^{\ell}_b$, consists of two layers of cells belonging to level $\ell$. In the inner layer (marked by circles), the density at the cell centers is linearly interpolated from the coarse level $\ell-1$.  In the outer layer (marked by crosses), the potential at the cell centers is linearly interpolated from the coarse level $\ell-1$ as well. The objective is to evaluate the self-gravitational potential for the gray region, i.e., $\Omega^{\ell}+$the first layer of the buffer zone, and obtain the self-gravitational forces in the $x$-direction at the cell centers of $\Omega^{\ell}$ using the fourth-order finite-difference:
\begin{eqnarray}
\label{Eqn:force_evaluation}
   \partial_x\Phi^{\ell} = \frac{4}{3}\frac{\Phi^{\ell}_{i+1}-\Phi^{\ell}_{i-1}}{2\Delta x^{\ell}}-\frac{1}{3}\frac{\Phi^{\ell}_{i+2}-\Phi^{\ell}_{i-2}}{4\Delta x^{\ell}}+\mathcal{O}((\Delta x^{\ell})^4),  
\end{eqnarray}
where we have used the subscript $i$ to index the cell centers in $x$ direction, and will use indices $j$, $k$ for $y$, $z$ directions in the following discussion. The evaluation of Eq.~(\ref{Eqn:force_evaluation}) involves information from the five nearest points in the direction of interest and this requirement sets the thickness of the buffer zones. 
\subsection{Build a multi-resolution hierarchy for a multigrid method}
\begin{figure}
\epsscale{1.0}
\plotone{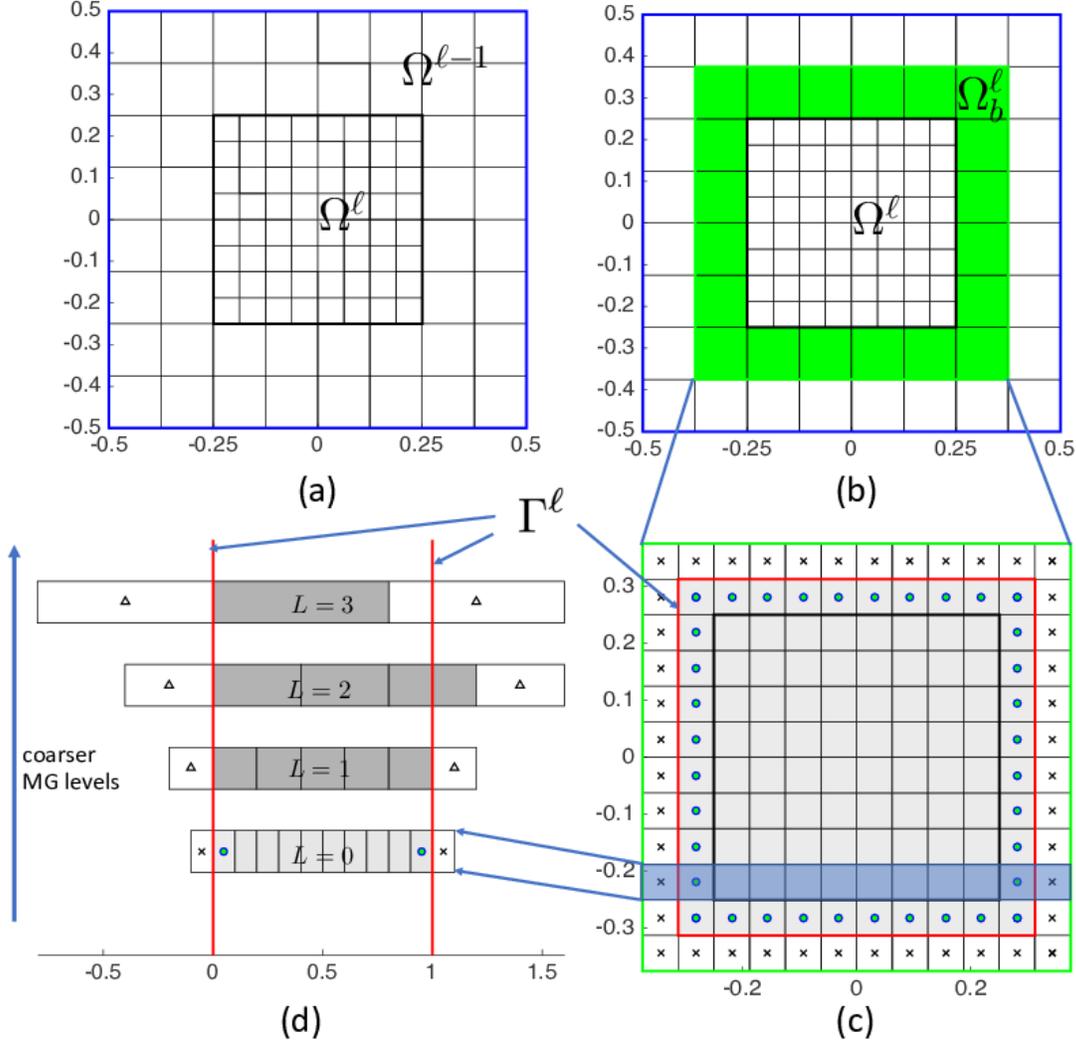}
\caption{(a) A two-dimensional representation of a refined region ($\Omega^{\ell}$) of level $\ell$ enclosed by the black rectangular. The domain of a coarse level $\ell-1$ enclosed by the blue rectangle is denoted as $\Omega^{\ell-1}$. A refined level should completely lies within the domain of a coarse level and the ratio of cell size between a coarse and a refined levels are fixed at two. (b) This refined domain $\Omega^{\ell}$ is surrounded by a buffer layer of two-cell size denoted as $\Omega^{\ell}_b$ (the green region). (c) The density values (denoted by circles) in the inner layer of $\Omega^{\ell}_b$ are linearly interpolated from the associated coarse region in $\Omega^{\ell-1}$. The potential values (denoted by crosses) in the outer layer of $\Omega^{\ell}_b$ are obtained by linear interpolation from the associated coarse region in $\Omega^{\ell-1}$. We apply the multigrid method to solve for the potential of the gay-shaded region enclosed by the red border denoted as \hhwang{$\Gamma^{\ell}$}. (d) A one-dimensional representation of the mesh structure used for the multigrid method. In this figure, we use capital $L$ to denote the level hierarchy of the multigrid algorithm. The values in $L=0$ level is taken directly from the blue-shaded region in (c). The width of the red border is normalized to $[0, 1]$. Except for the level $L=0$, the correction potential values at the \hhwang{$\Gamma^{\ell}$} is set to zero for levels $L>0$ in the relaxation process. The boundary values (denoted as triangles) for the multigrid levels $L>0$ are calculated by a linear extrapolation process detailed in \S~\ref{sec:BoundaryValueMG}.}
\label{fig:MGgrid}
\end{figure}

We have now set the stage for applying the multigrid method to a patch as defined in Fig.~\ref{fig:MGgrid}c for a level $\ell$. A multigrid method utilizes a set of multi-resolution mesh hierarchy to efficiently damp errors of large and small wavenumbers. Given the finest multigrid level, i.e., a patch of a refined level  shown in Fig.~\ref{fig:MGgrid}c, the multigrid mesh hierarchy in $x$ direction is shown in Fig.~\ref{fig:MGgrid}d, where the bottom-most one-dimensional mesh corresponds to the blue-shaded region in Fig.~\ref{fig:MGgrid}c. The multigrid hierarchy is built by a mesh coarsening process. This process goes from the bottom to the top, with the multigrid mesh sizes doubled with increasing multigrid levels until only one active cell (dark gray) is enclosed within the domain $\Omega^{\ell}$. Instead of using the multigrid boundary reconstruction algorithm proposed in \citet{gui11}, we introduce a simple algorithm that determines the number of active cells level-by-level as follows: 
\begin{enumerate}
\item Normalize the width of the light-gray region (\hhwang{$\Gamma^{\ell}$}) to a range $[0, 1]$ (including one buffer cell at the ends of either sides), and calculate the multigrid mesh size for the finest level $h^0_x \equiv 1.0/({\rm number~of~cells})$. The number of cells shown in Fig.~\ref{fig:MGgrid}d of $L=0$ is 10. 
\item Calculate the multigrid mesh size at level $L>0$ using the relation $h^L_x = 2h^{L-1}_x$.  
\item Calculate the number of active cells at the multigrid level $L>0$ by $N^{L}_{\rm MG} = {\rm round}(1/h^{L}_x)$. 
\item If $N^{L}_{\rm MG} > 1$, go to item 2; or stop, otherwise. 
\end{enumerate}
In the above and in the following discussion, we have used the superscript $L$ to denote the multigrid level to avoid possible confusion with mesh refinement level $\ell$. We note that in this paper the multigrid levels are described as a distinct hierarchical structure from the mesh refinement levels. While increasing $L$ goes from fine to coarse multigrid levels, increasing $\ell$ goes from coarse to fine refinement levels. This allows us to separate the details when describing the multigrid boundaries and refinement boundaries. Furthermore, the multigrid algorithm as described above can be applied directly to a uniform mesh, i.e., without AMR. The cell numbers in one-dimension can be arbitrary and do not have to be a power of two. For clarity, we also use the subscript MG to denote quantities associated with the multigrid relaxation process. In Fig.~\ref{fig:MGgrid}d, the left edges of all multigrid levels are aligned at $x_{\rm MG}=0$, here $x_{\rm MG}$ represents the normalized coordinates used for multigrid relaxation.  Given the $N^{L}_{\rm MG}$ and the $h^{L}_x$ obtained from the above algorithm, the right edges of active cells can be calculated, which are floating but always close to $x_{\rm MG}=1$. The active regions (marked by the dark gray) of all multigrid levels are then prepended and appended by a layer of buffer cells (marked by the triangles) for Dirichlet boundary conditions. 

\subsection{The multigrid iteration}\label{sec:multigrid_iteration}
A multigrid method involves a relaxation smoother and a multi-resolution hierarchy, which is built in the last subsection. Given the Dirichlet boundary conditions for all multigrid levels (will be detailed in the next subsection), we follow the conventional procedure for the multigrid iteration $L$ given a refinement level $\ell$:

\begin{enumerate}
\item[1:] {\bf do \{ }
	\begin{enumerate}
	\item[2:] $L=0$. Perform $N_{\rm pre}$ Gauss-Seidel smoothing iterations on the potential $\Phi^{L=0}_{\rm MG}$. 
	\item[3:] {\bf do \{ }
	\begin{enumerate}
		\item[4:] Compute the density residual: 
			\begin{eqnarray}
			\rho^{L}_{\rm res}= \nabla^2 \Phi^{L}_{\rm MG}-4\pi G\rho^{L}_{\rm MG}.
			\end{eqnarray}
		\item[5:] From fine to coarse levels, perform the restriction of the density residual
			\begin{eqnarray}
			\rho^{L+1}_{\rm MG}= \mathcal{R}(\rho^{L}_{\rm res}).
			\end{eqnarray}
		\item[6:] Perform $N_{\rm pre}$ smoothing iterations for the following problem:
			\begin{eqnarray}
			\nabla^2 \Phi^{L+1}_{MG} = - \rho^{L+1}_{\rm MG},
			\end{eqnarray}
			subject to the boundary condition:
			\begin{eqnarray}
			\Phi^{L+1}_{MG} = 0 \mbox{ on } \Gamma^{\ell}. 
			\end{eqnarray}
		\item[7:] $L \leftarrow L+1$.
	\end{enumerate}
	
	\item[8:] {\bf \} while} ($N^L_{\rm MG} > 1$)
	
	\item[9:] {\bf do \{}
		\begin{enumerate}
			\item[10:] Perform prolongation from $\Phi^{L}_{\rm MG}$ and do correction to  $\Phi^{L-1}_{\rm MG}$:
				\begin{eqnarray}
				\Phi^{L-1}_{\rm MG} \leftarrow \Phi^{L-1}_{\rm MG}+\mathcal{P}(\Phi^{L}_{\rm MG}). 
				\end{eqnarray}
			\item[11:] Perform $N_{\rm post}$ smoothing iterations on $\Phi^{L-1}_{\rm MG}$ for the following problem:
				\begin{eqnarray}
				\nabla^2 \Phi^{L-1}_{\rm MG} = - \rho^{L-1}_{\rm MG},
				\end{eqnarray}
				subject to the boundary condition:
				\begin{eqnarray}
				\Phi^{L-1}_{\rm MG} = 0 \mbox{ on } \Gamma^{\ell}. 
				\end{eqnarray}
			\item[12:] $L \leftarrow L-1$
			\end{enumerate}
	\item[13:] {\bf \} while} ($L>0$)
	
	\end{enumerate}
\item[14:] {\bf \} while} (Gravitational accelerations have not yet converged to a tolerable level.)
\end{enumerate}

In the above procedure, we need to specify the numbers of iteration $N_{\rm pre}$, $N_{\rm post}$ for smoothing. A restriction operator $\mathcal{R}$, a prolongation operator $\mathcal{P}$ for propagating information back and forth between levels and a discretization of the Laplace operator are described below. 

\noindent\begin{enumerate}
\item[1]{\it Discretization of the Laplace operator}
\end{enumerate}

For the Laplace operator in two dimensions, we adopt the five-point stencil finite difference approximation: 
\begin{eqnarray}
\label{eqn:Laplacian2D}
\nabla^2 (\Phi^{L}_{\rm MG})_{i,j} \approx \frac{1}{(\Delta x^{L}_{\rm MG})^2} [(\Phi^{L}_{\rm MG})_{i-1,j}+(\Phi^{L}_{\rm MG})_{i+1,j}+(\Phi^{L}_{\rm MG})_{i,j-1}+(\Phi^{L}_{\rm MG})_{i,j+1}-4(\Phi^{L}_{\rm MG})_{i,j}],
\end{eqnarray}
while for the three-dimensions, the seven-point stencil finite difference method reads:
\begin{eqnarray}
\label{eqn:Laplacian3D}
\nabla^2 (\Phi^{L}_{\rm MG})_{i,j,k} &\approx& \frac{1}{(\Delta x^{L}_{\rm MG})^2} [(\Phi^{L}_{\rm MG})_{i-1,j,k}+(\Phi^{L}_{\rm MG})_{i+1,j,k}+(\Phi^{L}_{\rm MG})_{i,j-1,k}+(\Phi^{L}_{\rm MG})_{i,j+1,k}+ \nonumber \\ 
&&(\Phi^{L}_{\rm MG})_{i,j,k-1}+(\Phi^{L}_{\rm MG})_{i,j,k+1}-6(\Phi^{L}_{\rm MG})_{i,j,k}],
\end{eqnarray}
These Laplace operators are second-order accurate. 

\noindent\begin{enumerate}
\item[2]{\it Restriction operator}
\end{enumerate}

\begin{figure}
\epsscale{1.0}
\plotone{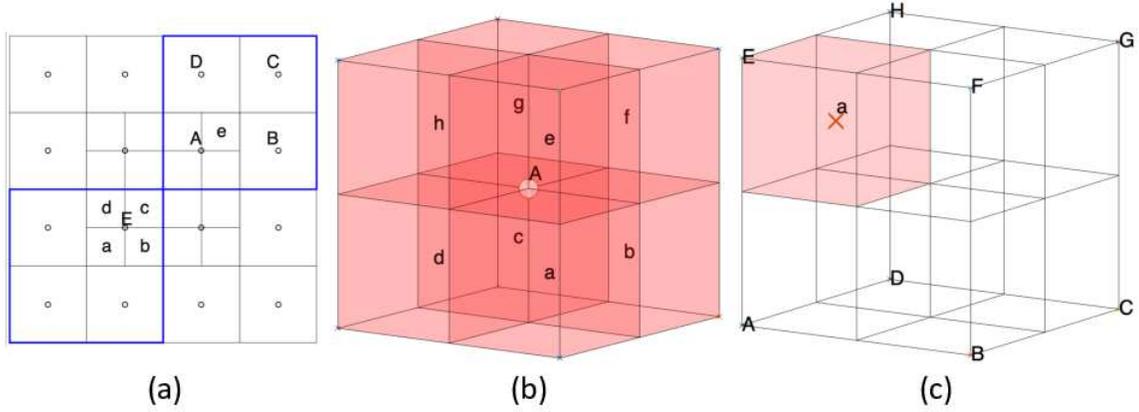}
\caption{These figures provide the visual realizations of the two- and three-dimensional restriction and prolongation operators. The upper-case letters denote the values at the cell centers of a coarser level, while the letters in lower-case are those values at the cell centers of a refined level. (a) The bottom-left blue rectangle shows the relative positions of a coarse and refined cells when applying the two-dimensional restriction operator $\mathcal{R}$ defined in Eq.~(\ref{eqn:restriction2D}). The upper-right blue rectangle shows the relative positions of coarse and a refined cells when applying the two-dimensional prolongation operator $\mathcal{P}$ defined in Eq.~(\ref{eqn:prolongation2D}). (b) The relative positions of a coarse and refined cells when applying the three-dimensional restriction operator $\mathcal{R}$ defined in Eq.~(\ref{eqn:restriction3D}). (c) The relative positions of coarse and a refined cells when applying the three-dimensional prolongation operator $\mathcal{P}$ defined in Eq.~(\ref{eqn:prolongation3D}). }
\label{fig:restriction_prolongation}
\end{figure}

For the restriction operator, we adopt the simple second-order scheme as shown in Fig.~\ref{fig:restriction_prolongation}. For the two-dimensional restriction operator, we consider the situation shown in the bottom-left blue square in Fig.~\ref{fig:restriction_prolongation}a, the value $E$ at the center of a coarse cell is evaluated by a simple average:
\begin{eqnarray}
\label{eqn:restriction2D}
E = \frac{1}{4}(a+b+c+d), 
\end{eqnarray}
where $a, b, c, d$ are values located at cell centers of the fine level. Similarly, for the three-dimensional restriction as shown in Fig.~\ref{fig:restriction_prolongation}b, the value $A$ at the center of a coarse cell is evaluated through:
\begin{eqnarray}
\label{eqn:restriction3D}
A = \frac{1}{8}(a+b+c+d+e+f+g+h),  
\end{eqnarray}
where $a, b, c, d,e,f,g,h$ are values located at cube centers of the fine level. The coefficients used in Eqs~(\ref{eqn:restriction2D})(\ref{eqn:restriction3D}) are derived in Appendix A.

\noindent\begin{enumerate}
\item[3]{\it Prolongation operator}
\end{enumerate}
For the prolongation operators, we adopt the second-order scheme as shown in Fig.~\ref{fig:restriction_prolongation}. This operator involves the nearest values from the coarse level that enclose the point of interest. For the two-dimensional prolongation as illustrated in the upper-right blue square in Fig.~\ref{fig:restriction_prolongation}a, the value of $e$ located at the cell center of the fine level is evaluated using the relation:
\begin{eqnarray}
\label{eqn:prolongation2D}
e = \frac{1}{16}(9A+3B+C+3D), 
\end{eqnarray}
where $A,B,C,D$ are values located at the cell centers of the coarse level. For the three-dimensional prolongation as shown in Fig.~\ref{fig:restriction_prolongation}c, the value $a$  at the center of the cube of fine level can be calculated with:
\begin{eqnarray}
\label{eqn:prolongation3D}
a = \frac{1}{64}(27E+9A+9H+9F+3G+3D+3B+C), 
\end{eqnarray}
where $A,B,C,D, E, F, G,H$ are values located at the cell centers of the coarse level. The coefficients used in Eqs~(\ref{eqn:prolongation2D})(\ref{eqn:prolongation3D}) are derived in Appendix A.

\subsection{Boundary conditions for multigrid}
\label{sec:BoundaryValueMG}
For a smoother to work, one needs to specify appropriate potential values to the boundary layers of all multigrid levels. For the finest multigrid level ($L=0$), the boundary potential (marked by crosses) can be interpolated from the potential calculated from the coarse level $\ell-1$. For the finest multigrid level ($L=0$), the density of the light-gray cells is taken directly from the blue-shaded region in Fig.~\ref{fig:MGgrid}c, while the initial potential can be interpolated from the potential calculated for the coarse level $\ell-1$. For the coarse multigrid levels ($L>0$), we require the coarse potential correction $\Phi^{L}_{\rm MG}=0$ at \hhwang{$\Gamma^{\ell}$}, i.e., steps 6 and 11 in \S\ref{sec:multigrid_iteration}. This requirement gives a simple way to specify the boundary values for multigrid coarse level $L>0$. For simplicity, we consider the one-dimensional example as shown in Fig.~\ref{fig:MGgrid}d. For the left boundaries, we simply adopt $(\Phi^{L}_{MG})_{0} = -(\Phi^{L}_{MG})_{1}$. Here, we use the subscript $i=1, 2, ..., N^{L}_{\rm MG}$ in $(\Phi^{L}_{MG})_{i}$ to denote the cell centers of active cells (dark gray cells), and $i=0$ and $i=N^{L}_{\rm MG}+1$ for the cell centers of left and right buffer zones, respectively. For the values in right buffer zones, we extrapolate using the following rules:
 \begin{eqnarray} \label{eqn:boundary_value}
 (\Phi^{L}_{MG})_{N^{L}_{MG}+1} = \left\{
\begin{array}{cc}
-(\Phi^{L}_{MG})_{N^{L}_{MG}-1}, & \mbox{ if } (x^{L}_{\rm MG})_{N^{L}_{\rm MG}} = 1 \\
(\Phi^{L}_{MG})_{N^{L}_{MG}}\left(\frac{1.0-h^{L}_x}{1.0-(x^{L}_{\rm MG})_{N^{L}_{\rm MG}}}\right) ,& \mbox{ otherwise} 
\end{array}
\right., 
 \end{eqnarray}
where $(x^{L}_{\rm MG})_{N^{L}_{\rm MG}} $ is the center of the right-most dark-gray cell of level $L$. Equation~(\ref{eqn:boundary_value}) shall apply to $y$ and $z$ boundaries, respectively, depending on the dimension of a problem. We note that for every multigrid iteration given a coarse multigrid level $L$, the boundary value should be updated according to the updated $\Phi^{L}_{\rm MG}$ in order to satisfy the condition $\Phi^{L}_{\rm MG}=0$ on \hhwang{$\Gamma^{\ell}$}. 

\section{simulations and Results}
In this section, we apply the multigrid procedure detailed in \S~\ref{sec:method} to a two- and a three-dimensional problems. The examples are carefully selected so that the analytic expression of forces is sufficiently smooth for probing the order of a numerical method. In order to quantify the rate of convergence, we measure the errors between numerical and analytic solutions using the $\|\cdot\|_{p}$-norm defined as:
\begin{eqnarray}
\| \mathcal{F}\|_{p} = \left(\int_{\Omega}|\mathcal{F}(\vec{x})|^p {\rm d}\vec{x}\right)^{1/p}, \mbox{  if } p\ge 1,
\end{eqnarray}
and 
\begin{eqnarray}
\|\mathcal{F}\|_{\infty} = \mbox{sup}_{\Omega}|\mathcal{F}(\vec{x})|, \mbox{  if } p\rightarrow \infty,
\end{eqnarray}
where $\Omega$ represents the entire calculation domain. In the following discussion, we measure the convergence of the proposed method using the one-, two- and infinity-norms. When using $\|\cdot\|_{1}$ and $\|\cdot\|_{2}$, we evaluate the total variation and energy in errors, while using $\|\cdot\|_{\infty}$ we monitor the convergence of maximum errors in a uniform sense. Here, ``uniform" means that {\it all} the errors in the calculation domain approach zero at the {\it same} order of converging speed. Finally, for the following tests, we use $N_{\rm pre}=2$ and $N_{\rm post}=1$ for the best performance. 

\subsection{A 2D model}\label{2Dmodel}
For the two-dimensional test, we solve a Poisson equation with $G=1/(4\pi)$:
\begin{eqnarray}
\nabla^2\Phi_{\rm 2D} = \sigma,
\end{eqnarray}
where $\Phi_{\rm 2D}$ is the 2D potential and the density $\sigma$ has the following distribution:
\begin{eqnarray}
\sigma(R) = \frac{4 R_0^2}{(R^2+R_0^2)^2}, 
\end{eqnarray}
where $R\equiv \sqrt{x^2+y^2}$ is the polar radius. The corresponding analytic potential reads:
\begin{eqnarray}
\label{Eq:PhiR}
\Phi_{\rm 2D}(R)=\ln\left[ \left( \frac{R}{R_0}\right)^2+1 \right],  
\end{eqnarray}
and the analytical form of the gravitational accelerations in radial and $x$ directions are $f_R\equiv |\nabla \Phi_{\rm 2D}|$ and $f_x$:
\begin{eqnarray}
\label{Eq:fR}
f_R=\frac{2R}{R^2+R_0^2}, \\
f_x=f_R\frac{x}{R}. 
\end{eqnarray}

The calculation domain of level $\ell$ is defined in Cartesian coordinates $\Omega^{\ell} = [\omega_L^{\ell},\omega_R^{\ell}]\times[\omega_F^{\ell},\omega_B^{\ell}]$, with $\ell=0$ represents the base level. For the current example, we use $[\omega_L^0, \omega_R^0]\times [\omega_F^0, \omega_B^0] = [-0.5, 0.5]\times [-0.5, 0.5]$, and $\Omega^{\ell} = [\omega_L^{\ell-1}/2, \omega_R^{\ell-1}/2] \times [\omega_F^{\ell-1}/2, \omega_B^{\ell-1}/2]$ for $\ell > 0$. Three levels of refinement are applied in this example, i.e., ${\ell} = 0,1,2,3$. The simulation is repeated for different cell numbers, $N=8,16,32,64,128,256,512,1024$, in one direction of the base level ($\ell=0$). Since the calculation domain in one direction of level $\ell$ also shrinks half compared to level $\ell-1$, we have $N^{\ell} = N$. In this example, we set $R_0=0.3$. If we count one round of the procedure described in \S\ref{sec:multigrid_iteration} as one iteration, it normally takes around 10 or fewer iterations to converge the numerical results to a level ${\rm max_{\Omega}}|f_{R,{\rm num}}^{q}-f_{R, {\rm num}}^{q-1}| < 10^{-10}$, where $f_{R, {\rm num}}^{q}$ represents the numerical radial acceleration obtained at the $q$th iteration. The iteration number required for convergence is independent of $N$.

We define the absolute differences between the numerical results and the analytic expressions:  
\begin{eqnarray}
{L}_R(\vec{x}) = |f_{R, {\rm num}}-f_{R}|(|\vec{x}|), \\
{L}_x(\vec{x}) = |f_{x, {\rm num}}-f_{x}|(|\vec{x}|), 
\end{eqnarray}
where $\vec{x}$ represents $(x,y)$ in 2D and $(x,y,z)$ in 3D. We apply these expressions to measure their one, two and infinity error norms. The results are listed in Table~\ref{tbl:2D_model}. The upper table tabulates the numerical values of error norms with increasing $N$, while the lower table tabulates the orders of accuracy improvement when the $N$ of the base level is doubled. This Table shows that the multigrid method proposed in this work is of second-order accuracy in terms of $L^1$, $L^2$ and $L^{\infty}$ norms. 

\begin{figure}
\epsscale{1.0}
\plotone{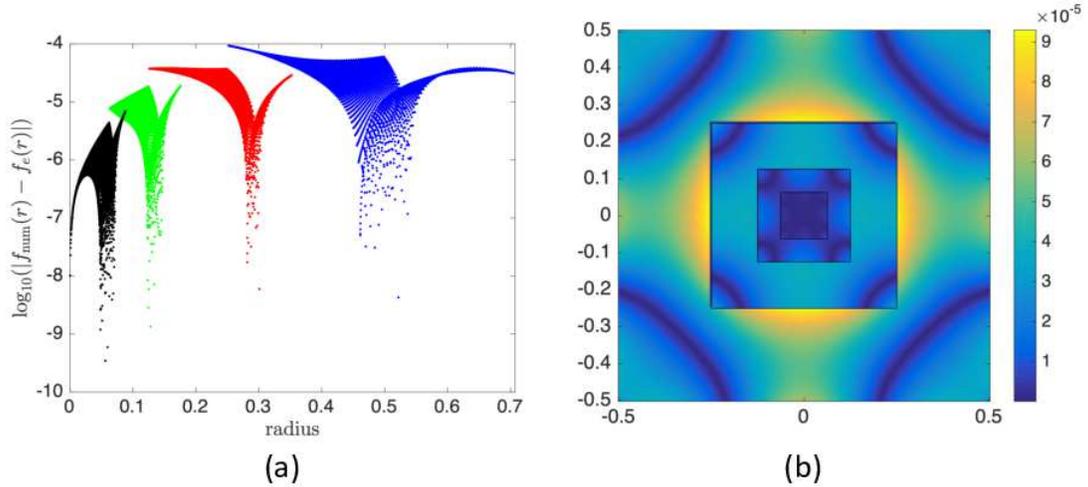}
\caption{(a) The errors of radial acceleration as a function radius. Each dot of different color corresponds to the result of one cell of different refined levels. Spurious forces are largely suppressed at the interfaces between coarse and refined levels (cf. Fig.~11 in \citet{gui11}). (b) The map of acceleration errors. }
\label{fig:fr_error_map}
\end{figure}

Figure~\ref{fig:fr_error_map}a shows the $L_R(\vec{x})$ as a function of $R$ for $N=256$. The blue, red, green and black dots correspond to cells of levels $\ell=0,1,2,3$, respectively. The spurious forces at the interfaces between levels are effectively suppressed. In Fig.~\ref{fig:fr_error_map}b, the two-dimensional error map shows that the numerical errors are still subject to the "squareness" of the Cartesian domain, however, the errors are steadily decreasing with increasing level of refinement. 

\subsection{A 3D model}
For the three-dimensional test, we solve for a Poisson equation with $G=1$:
\begin{eqnarray}
\nabla^2\Phi_{3D} =4\pi\rho.
\end{eqnarray}
The volume density $\rho$ has the form:
 \begin{eqnarray}
 \rho(r)\equiv\left\{
\begin{array}{cc}
\rho_0\left(1- r^2/r_0^2\right)^n, & \mbox{ if } r \le r_0 \\
0 ,& \mbox{ if } r  > r_0
\end{array}
\right. ,
 \end{eqnarray}
where $n=2$ and $\rho_0=1$ are used for the current work, $r\equiv \sqrt{x^2+y^2+z^2}$ is the spherical radius and $r_0=0.25$ is a parameter controlling the size of the ball. Imposing the boundary condition $\lim_{r \rightarrow \infty}\Phi(r) \rightarrow 0$, the analytic expression for the potential reads:
 \begin{eqnarray}
 \Phi(r)_{3D}\equiv\left\{
\begin{array}{cc}
-\frac{2}{3}\pi\rho_0 r_0^2+4\pi\rho_0(\frac{r^2}{6}-\frac{1}{10}\frac{r^4}{r_0^2}+\frac{1}{42}\frac{r^6}{r_0^4}), & \mbox{ if } r \le r_0 \\
-M_{\mbox{\rm ball}}/r,& \mbox{ if } r  > r_0
\end{array}
\right. ,
 \end{eqnarray}
 where $M_{\mbox{\rm ball}}=32\pi \rho_0 r_0^3/105$ is the total mass of the ball. The corresponding expression for the radial acceleration $f_r = - \nabla \Phi$ then reads:
 \begin{eqnarray}
 f_r(r)\equiv\left\{
\begin{array}{cc}
-4\pi\rho_0(\frac{r}{3}-\frac{2}{5}\frac{r^3}{r_0^2}+\frac{1}{7}\frac{r^5}{r^4_0}), & \mbox{ if } r \le r_0 \\
-M_{\mbox{\rm ball}}/r^2,& \mbox{ if } r  > r_0
\end{array}
\right., 
 \end{eqnarray}
 and $f_x(r) = f_r(r) (x/r)$. 
 The force expressions are sufficiently smooth for validating a method of second order accuracy. 

For the numerical setup, the calculation domain of level $\ell$ is defined in Cartesian coordinates $\Omega^{\ell} = [\omega_L^{\ell},\omega_R^{\ell}]\times [\omega_F^{\ell}, \omega_B^{\ell}]\times [\omega_D^{\ell}, \omega_T^{\ell}]$, with $\ell=0$ denotes the base level. For the base level, $\Omega^0=[-0.5,0.5]\times [-0.5, 0.5]\times [-0.5, 0.5]$ and $\Omega^{\ell} = [\omega_L^{\ell-1}/2,\omega_R^{\ell-1}/2]\times [\omega_F^{\ell-1}/2, \omega_B^{\ell-1}/2]\times [\omega_D^{\ell-1}/2, \omega_T^{\ell-1}/2]$ for $\ell > 0$. Two levels of refinement are applied in this example, i.e., ${\ell} = 0,1,2$. The simulation is performed with cell numbers $N=8,16,32,64,128,256$, in one direction of the base level ($\ell=0$). Similar to the two-dimensional model, if the initial guess of $\Phi^{\ell}$ is interpolated from a coarse level $\Phi^{\ell-1}$, it takes less than 10 iterations to converge the numerical results to a level ${\rm max_{\Omega}}|f_{r,{\rm num}}^{q}-f_{r, {\rm num}}^{q-1}| < 10^{-10}$, where $f_{r, {\rm num}}^{q}$ represents the numerical radial acceleration obtained at the $q$th iteration. The iteration number required for convergence is independent of $N$.

We apply the $L^1$, $L^2$ and $L^{\infty}$ norms to measure the acceleration errors in radial and $x$ directions and the results are listed in Table~\ref{tbl:3D_model}. The interpretation of this table is the same as Table~\ref{tbl:2D_model}. The results confirm that the proposed multigrid algorithm is of second-order accuracy in terms of all error norms. The spurious forces at the interfaces between different levels are suppressed in a three-dimensional problem. The proposed algorithm is generally applicable to nested mesh refinement in Cartesian coordinates. 

\section{Summary and Discussions}

Based on the previous work of \citet{gui11}, we present an improved multigrid Poisson solver of second-order accuracy in terms of all error norms. The method is applicable to grid-based HD/MHD codes which adopt patch-based mesh refinement schemes. The spurious forces commonly seen at interfaces between coarse and fine levels are largely suppressed so that the accuracy of self-gravitational accelerations derived from the potential also reach second-order accuracy in terms of all error norms. 

Compared to the work of \citet{gui11}, we have the following major differences in terms of algorithm:
\begin{enumerate}
\item A buffer zones of two-cell width is required. Although the algorithm proposed by \citet{gui11} requires only one layer of ghost cells, this modification shall not be considered as a major drawback of the method given the  improvement in accuracy. 
\item The proposed algorithm does not need a mask function for boundary reconstruction in the process of multigrid relaxation.
\item The proposed algorithm does not need a modified Laplace operator when approaching the boundary of a patch. One may apply Eqs.~(\ref{eqn:Laplacian2D}) and ~(\ref{eqn:Laplacian3D}) uniformly over the domain $\Omega^{\ell}\cup\Omega^{\ell}_b$. 
\end{enumerate}
Overall, the proposed algorithm is simpler, easier to implement and more accurate. 

\begin{figure}
\epsscale{0.8}
\plotone{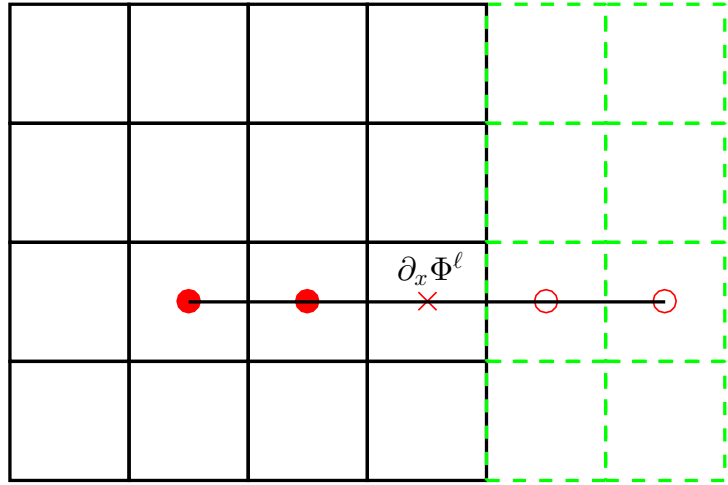}
\caption{Calculation of gravitational forces at the fine-coarse boundary using Eq.~(\ref{Eqn:force_evaluation}). Evaluating the gravitational force for the cell (marked by the cross) next to the boundary involves fives potentials marked with the solid circles, the cross, and the empty circles. The solid circles and the cross represent the potential of the fine level, while the empty circles are the potential in the buffer layer. While \citet{gui11} computed the values of empty circles from the coarse level potential by linear interpolation, in this work, the potential in the buffer layer are directly calculated in the process of relaxation to avoid a jump in truncation errors across the fine-coarse boundary.}
\label{fig:boundary_derivative}
\end{figure}

We attribute the second-order accuracy in derivatives at refinement boundaries to the use of buffer layers of two-cell size. As shown in Fig.~\ref{fig:boundary_derivative},  evaluating Eq.~(\ref{Eqn:force_evaluation}) for the cell (shown as the red cross) adjacent to the boundary involves five cells as indicated by the solid circles, the cross and the empty circles. The empty circles represent the potentials in the buffer zones. \citet{gui11} fill the values in the buffer zones by linear interpolation from the coarser levels. As already remarked in their work, since the coarse Laplace operator and the linear interpolation are both accurate to second order, the truncation errors associated with the interpolated values (empty circles) have the form $(\Delta x)^2\epsilon_i$, while those associated with the fine level (solid circles) have $(\Delta x)^2\eta_i$, where $\epsilon_i$ and $\eta_i$ are of the order $\mathcal{O}(1)$. Since the potential of different levels are calculated separately, $\epsilon_i$ does not smoothly connect to $\eta_i$ in general. When applying Eq.~(\ref{Eqn:force_evaluation}), the jump in second order truncation errors degrades the accuracy to first order across fine-coarse boundaries. Without resorting to higher order Laplace and interpolation operators, in this work, the values in the buffer layers are evaluated directly in the relaxation process of fine levels. This enforces a smooth connection of truncation errors across boundaries, therefore the second-order accuracy is maintained.

Finally, we note that the multigrid relaxation method is fast and flexible, especially when mesh refinement is required. For the self-gravitational forces of infinitesimally thin disks, using multigrid method will not benefit much since an appropriate boundary conditions, which are also unknown, need to be calculated in advance and imposed on the surface of a thin box, thus making the method complicated and inefficient. Recently, \citet{moo19} combine the James algorithm \citep{jam77} with the discrete Green's function (DGF) to develop an accurate and efficient algorithm that may apply to three-dimensional Cartesian and cylindrical coordinates with open boundary conditions. This algorithm has a complexity of order $\mathcal{O}(N^3\log N)$, with $N$ being the number of cells in one-dimension. In their work, the evaluation of boundary potentials using DGF is the key to reach second-order accuracy in potential. The use of DGF is interesting and may combine with the multigrid method for mesh refinement. We leave explorations along this line for future works. Direct, fast and accurate methods that reduce three-dimensional infinitesimally disk problems to two-dimensional integrals are also developed. The kernel-based Poisson solvers of high-order accuracy for self-gravitational forces for infinitesimally thin disks have been extensively studied for uniform Cartesian coordiantes \citep{yen12, wang19}, for nested mesh refinement in Cartesian coordinates \citep{wang16}, for adaptive mesh refinement using GPU acceleration \citep{tse19}, and for polar coordinates \citep{wang15}. 

\acknowledgements
C.C.Y. thanks the Institute of Astronomy and Astrophysics, Academia Sinica, Taiwan for their constant support. C.C.Y. is supported by Ministry of Science and Technology of Taiwan, under grant MOST-107-2115-M- 030-005-MY2. HHW thanks the support by the grant from the Research Grants Council of Hong Kong: General Research Fund 14308217, 14305717 and the two supports by the Research Committee Direct Grant for Research from CUHK: 4053229, 4053309.


\begin{table}[ht]
\begin{center}
\begin{tabular}{|c|c|c|c|c|c|c|}
\hline
$N$ & $L_{x}^1$ & $L_{x}^2$ & $L_{x}^{\infty}$ & $L_{R}^1$ & $L_{R}^2$ & $L_{R}^{\infty}$ \\ \hline
8   &  1.878e-2 &   2.528e-2& 7.181e-2  & 2.904e-2 & 3.549e-2 & 7.261e-2 \\ \hline
16   & 4.628e-3 &  6.232e-3 & 2.002e-2 & 7.133e-3 &  8.764e-2& 2.024e-2 \\ \hline
32   &  1.186e-3&   1.601e-3&  5.444e-3 &  1.847e-3&  2.244e-3& 5.463e-3 \\ \hline
64   &  3.100e-4 &  4.147e-4& 1.430e-3  & 4.793e-4&  5.774e-4& 1.432e-3\\ \hline
128  &  8.016e-5 &  1.065e-4& 3.671e-4  & 1.229e-4& 1.474e-4& 3.672e-4\\ \hline
256 &  2.045e-5&   2.705e-5&  9.303e-5 & 3.119e-5&  3.731e-5& 9.303e-5\\ \hline
512 &  5.168e-6&  6.821e-6 & 2.342e-5   &  7.860e-6&  9.389e-6& 2.342e-5\\ \hline
1024   &  1.299e-6 &  1.713e-6& 5.874e-6 &  1.973e-6& 2.356e-6&  5.874e-6\\ \hline
\hline
$N_p/N_{p+1}$ &  $O_{x}^1$ & $O_{x}^2$ & $O_{x}^{\infty}$ & $O_{R}^1$ & $O_{R}^2$ & $O_{R}^{\infty}$ \\ \hline
8/16           & 2.02  &  2.02 & 1.84  & 2.03 &  2.02 &  1.84\\ \hline
16/32           & 1.96  &  1.96 &  1.88 & 1.95 &  1.97 &  1.89\\ \hline
32/64           &  1.94 &  1.95 &  1.93 &  1.95&  1.96 & 1.93 \\ \hline
64/128           &  1.95 &   1.96&  1.96 & 1.96 &  1.97 & 1.96 \\ \hline
128/256           & 1.97  &  1.98 &  1.98 & 1.98 &  1.98 & 1.98 \\ \hline
256/512           &  1.98 &  1.99 &  1.99 &  1.99&   1.99&  1.99\\ \hline
512/1024          &  1.99 &  1.99 &  2.00 & 1.99 &  1.99 & 2.00 \\ \hline
\end{tabular}
\caption{This table tabulates the errors and orders of accuracy of the $x$ and radial accelerations for the two-dimensional model. Three levels of refinement are applied. The subscript $p$ is used to indicate the comparisons between different cell numbers. For example, when $(N_p/N_{p+1})=(32/64)$, the $O^{1}_x = \log_2(1.186{\rm e-}3/3.100{\rm e-}4)\approx 1.94$.}  \label{tbl:2D_model}
\end{center}
\end{table}

\begin{table}[ht]
\begin{center}
\begin{tabular}{|c|c|c|c|c|c|c|}
\hline
$N$ & $L_{x}^1$ & $L_{x}^2$ & $L_{x}^{\infty}$ & $L_{r}^1$ & $L_{r}^2$ & $L_{r}^{\infty}$ \\ \hline
8   & 1.275e-3  &  2.191e-3    & 9.813e-3 & 1.849e-3& 3.386e-3 & 1.027e-2\\ \hline
16   & 3.255e-4  & 5.035e-4  & 3.390e-3 & 4.219e-4& 7.032e-4 & 3.467e-3\\ \hline
32  & 7.883e-5  & 1.253e-4  & 8.738e-4 & 9.702e-5& 1.678e-4 & 9.260e-4\\ \hline
64  & 1.958e-5  & 3.169e-5    & 2.159e-4 & 2.415e-5& 4.170e-5& 2.392e-4\\ \hline
128  & 4.867e-6  & 7.991e-6  &  5.340e-5& 6.011e-6& 1.040e-5& 6.073e-5\\ \hline
256  & 1.215e-6  & 2.010e-6  &  1.327e-5 & 1.505e-6& 2.605e-6& 1.530e-5\\ \hline
\hline
$N_p/N_{p+1}$ &  $O_{x}^1$ & $O_{x}^2$ & $O_{x}^{\infty}$ & $O_{r}^1$ & $O_{r}^2$ & $O_{r}^{\infty}$ \\ \hline
8/16           & 1.97& 2.12 & 1.53&  2.13&  2.27 &  1.57\\ \hline
16/32           & 2.05& 2.01 & 1.96&  2.12& 2.07  &  1.90\\ \hline
32/64           &  2.01&  1.98& 2.02& 2.01 & 2.01  &  1.95\\ \hline
64/128           & 2.01&  1.99& 2.02& 2.01 & 2.00  &  1.98\\ \hline
128/256          &2.00 &  1.99& 2.01& 2.00 &  2.00 &  1.99\\ \hline

\end{tabular}
\caption{This table tabulates the errors and orders of accuracy of the $x$ and radial accelerations for the three-dimensional model. Two levels of refinement are applied. The subscript $p$ is used to indicate the comparisons between different cell numbers. The subscript $p$ is used to indicate the comparisons between different cell numbers. For example, when $(N_p/N_{p+1})=(32/64)$, the $O^{1}_x = \log_2(7.883{\rm e-}5/1.958{\rm e-}4)\approx 2.01$.}  \label{tbl:3D_model}
\end{center}
\end{table}

\appendix
\section{bilinear and trilinear interpolations}
\begin{figure}
\epsscale{1.0}
\plotone{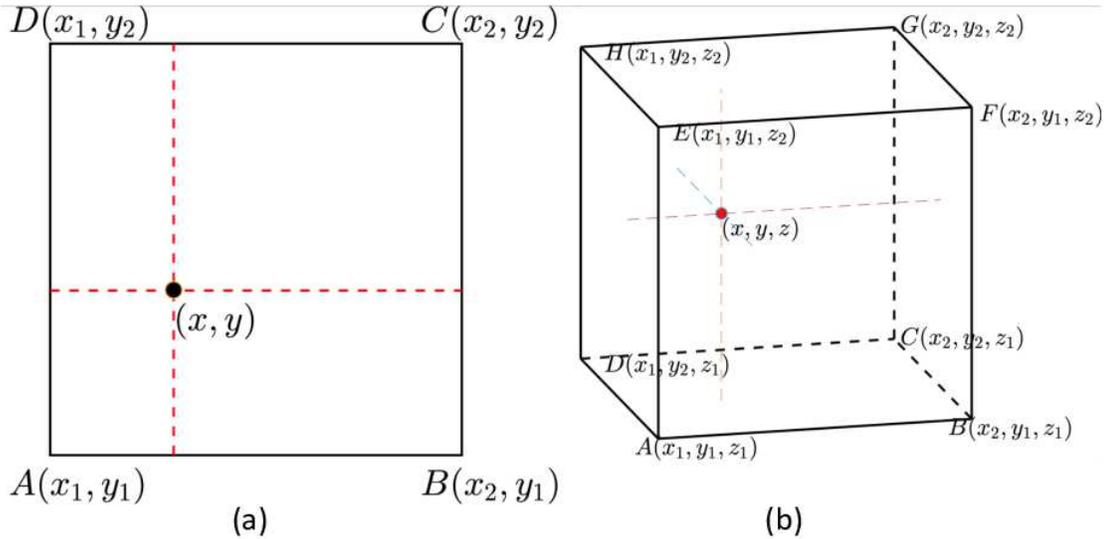}
\caption{(a) Bilinear interpolation. Given the values $A, B, C, D$ located at $(x_1,y_1)$, $(x_2,y_1)$, $(x_2,y_2)$, $(x_1,y_2)$, respectively, we apply the bilinear interpolation Eq.~(A1) to approximate the value located at $(x, y)$. (b) Trilinear interpolation. Given the values $A, B, C, D, E, F, G, H$ located at $(x_1,y_1, z_1)$, $(x_2,y_1, z_1)$, $(x_2,y_2, z_1)$, $(x_1,y_2, z_1)$, $(x_1,y_1, z_2)$, $(x_2,y_1, z_2)$, $(x_2,y_2, z_2)$, $(x_1,y_2, z_2)$, respectively, the trilinear interpolation Eq.~(A2) is applied to approximate the value located at $(x,y,z)$.}
\label{fig:interpolation_2D3D}
\end{figure}

In this section, we explain the origin of the coefficients used for the restriction operators Eqs.~(\ref{eqn:restriction2D})(\ref{eqn:restriction3D}), and the prolongation operators Eqs.~(\ref{eqn:prolongation2D})(\ref{eqn:prolongation3D}).  As shown in Fig.~\ref{fig:interpolation_2D3D}a, given the values located at positions $A(x_1,y_1), B(x_2,y_1), C(x_2,y_2), D(x_1,y_2)$, we approximate the value $f(x,y)$ using the bilinear interpolation $\tilde{f}(x,y)$:
\begin{eqnarray}
\tilde{f}(x,y) &=& \frac{1}{(\Delta x)(\Delta y)}[A(x_2-x)(y_2-y)+B(x-x_1)(y_2-y) \nonumber \\
    &&+C(x-x_1)(y-y_1)+D(x_2-x)(y-y_1)], 
\end{eqnarray}
where $\Delta x \equiv x_2-x_1$ and $\Delta y \equiv y_2-y_1$. Due to the squareness of the cell shape, we further impose the condition $\Delta y = \Delta x$. One may easily check that $\tilde{f}(x_1,y_1)=A$, $\tilde{f}(x_2,y_1)=B$, $\tilde{f}(x_2,y_2)=C$ and $\tilde{f}(x_1,y_2)=D$. Assume the function $f(x,y)$ varies smoothly over the domain $[x_1, x_2]\times [y_1, y_2]$, the interpolation $\tilde{f}$ approximates $f$ to an accuracy of second-order. Evaluating the value at the center where we have $x-x_1 = x_2-x = 0.5\Delta x$ and $y-y_1 = y_2-y = 0.5\Delta x$ naturally leads to the coefficients of Eq.~(\ref{eqn:restriction2D}). To consider the situation in the upper-right of Fig.~2a, one may simply take $x-x_1=0.25\Delta x$, $x_2-x = 0.75 \Delta x$, $y-y_1=0.25 \Delta y$ and $y_2-y = 0.75 \Delta y$, resulting in the coefficients in Eq.~(\ref{eqn:prolongation2D}). 

By the same token, as shown in Fig.~\ref{fig:interpolation_2D3D}b, the three-dimensional equivalent of Eq.~(A1) is the trilinear interpolation:
\begin{eqnarray}
\tilde{f}(x,y,z) &=& \frac{1}{(\Delta x)(\Delta y)}[A(x_2-x)(y_2-y)(z_2-z)+B(x-x_1)(y_2-y)(z_2-z) \nonumber \\
    &&+C(x-x_1)(y-y_1)(z_2-z)+D(x_2-x)(y-y_1)(z_2-z) \nonumber \\
    &&+E(x_2-x)(y_2-y)(z-z_1)+F(x-x_1)(y_2-y)(z-z_1) \nonumber \\
    &&+G(x-x_1)(y-y_1)(z-z_1)+H(x_2-x)(y-y_1)(z-z_1)].
\end{eqnarray}
The coefficients of Eq.~(\ref{eqn:restriction3D}) is then obtained using the conditions $x_2-x=x-x_1=y_2-y=y-y_1=z_2-z=z-z_1=0.5\Delta x$. Equation~(\ref{eqn:prolongation3D}) is a result of $x-x_1=y-y_1=z_2-z=0.25\Delta x$ and $x_2-x=y_2-y=z-z_1=0.75\Delta x$.


\bibliography{MG_gravity}

\end{document}